\documentclass[floats,prd,aps,amssymb,nofootinbib]{revtex4}
\usepackage{amssymb}
\usepackage{graphics}
\usepackage{graphicx}
\usepackage{amsmath}
\usepackage{color}
\usepackage{amsfonts}
\usepackage{bm}
\usepackage[]{latexsym}
\usepackage{float}

\newcommand{\be}{\begin{equation}}\newcommand{\ee}{\end{equation}}
\newcommand{\bea}{\begin{eqnarray}}\newcommand{\eea}{\end{eqnarray}}
\newcommand{\brr}{\begin{array}}\newcommand{\err}{\end{array}}
\newcommand{\bit}{\begin{itemize}}\newcommand{\eit}{\end{itemize}}
\newcommand{\ben}{\begin{enumerate}}\newcommand{\een}{\end{enumerate}}

\newcommand{\ba}{\begin{array}}
\newcommand{\ea}{\end{array}}

\def\lf{\left}

\def\non{\nonumber}\def\pa{\partial}

\def\ri{\right}
\def\al{\alpha}
\def\de{\delta}

\def\si{\sigma}

\def\1{{_{1}}}\def\2{{_{2}}}

\begin{document}
\title{Flavor vacuum entanglement in boson mixing}

\author{Massimo Blasone\footnote{blasone@sa.infn.it}$^{\hspace{0.3mm}1,2}$, Fabrizio Illuminati\footnote{filluminati@unisa.it}$^{\hspace{0.3mm}2,3}$, Giuseppe Gaetano Luciano\footnote{gluciano@sa.infn.it}$^{\hspace{0.3mm}1,2}$ and Luciano Petruzziello\footnote{lupetruzziello@unisa.it}$^{\hspace{0.3mm}2,3}$} \affiliation
{$^1$Dipartimento di Fisica, Universit\`a degli Studi di Salerno, Via Giovanni Paolo II, 132 I-84084 Fisciano (SA), Italy.\\ $^2$INFN, Sezione di Napoli, Gruppo collegato di Salerno, Italy. 
\\ $^3$Dipartimento di Ingegneria Industriale, Universit\`a degli Studi di Salerno, Via Giovanni Paolo II, 132 I-84084 Fisciano (SA), Italy.}

\date{\today}
\def\be{\begin{equation}}
\def\ee{\end{equation}}
\def\al{\alpha}
\def\bea{\begin{eqnarray}}
\def\eea{\end{eqnarray}}

\begin{abstract}
Mixing transformations in quantum field theory
are non-trivial, since they are intimately related to the 
unitary inequivalence between Fock spaces for fields
with definite mass and fields with definite flavor. Considering the 
superposition of two neutral scalar (spin-0) bosonic fields,
we investigate some features of the emerging condensate 
structure of the flavor vacuum. In particular, 
we quantify the flavor vacuum entanglement in terms of the 
von Neumann entanglement entropy of the reduced state. 
Furthermore, in a suitable limit, we show that the flavor vacuum 
has a structure akin to the thermal vacuum of 
Thermo Field Dynamics, with a temperature dependent 
on both the mixing angle and the particle mass difference.
\end{abstract}

\vskip -1.0 truecm 

\maketitle

\section{Introduction}

The analysis of quantum correlations in the context of particle physics (and in particular for neutrino and meson systems) is currently gaining a rising attention. The initial observation that flavor mixing and oscillations can be associated with (single-particle) entanglement~\cite{entmix,entmix2} served as a basis for several studies~\cite{numecorr,numecorr2,numecorr3,numecorr4,numecorr5,numecorr6,numecorr7,numecorr8,numecorr9} in which violations of Bell, Leggett-Garg and Mermin-Svetchlichny inequalities, non-locality, gravity/acceleration degradation effects and other similar occurrences have been investigated both theoretically and experimentally. 

Most of the above studies have been carried out within the framework of Quantum Mechanics (QM). The extension to Quantum Field Theory (QFT) has been later considered in Ref.~\cite{entQFT}, thus leading to the discovery of non-trivial properties of the mixing transformation~\cite{annals}. Indeed, whilst in QM such a 
transformation acts as a simple rotation between flavor and 
mass states~\cite{pontecorvo}, 
its QFT counterpart behaves as 
a rotation nested into a non-commuting 
Bogoliubov transformation~\cite{annals, bogoliubov}. 
As a result, the vacua for fields with definite
mass and fields with definite flavor become orthogonal to each
other, with the latter acquiring the 
structure of a $SU(2)$ coherent state~\cite{perelomov}
and turning into a condensate of massive particle/antiparticle pairs~\cite{annals,smald}. 
This gives rise to a deeeper understanding of particle mixing, since the Fock spaces for flavor and mass fields  
are found to describe {\it unitarily inequivalent} representations. 
Corrections to the standard QM predictions 
also appear in the oscillation probability formula, as explicitly shown in Ref.~\cite{exactoscill}. 

Originally developed for neutrinos propagating in flat spacetime, the above considerations have been later extended to bosonic fields~\cite{bosonmix1,bosonmix2} 
as well as to non-trivial spacetime backgrounds (see Refs.~\cite{prd,Capelup} for more details).  
Recently, a further evidence for the 
complex structure of the flavor vacuum
has been provided in Ref.~\cite{CUBA}, 
where it has been established that the Fock space
for flavor fields cannot be obtained by the 
direct product of the spaces for massive fields. 
Therefore, entanglement connected with flavor mixing appears to be an exquisite concept, boiling down to the non-factorizability of the flavor states in terms of the ones with definite masses. In other words, it is possible to come across flavor entanglement already at the level of the vacuum state. More generally, the phenomenon of a non-vanishing entanglement 
even for free fields is ultimately related to the quotient space structure of the tensor Hilbert spaces~\cite{casini}.

In fact, the investigation of the properties of the QFT vacuum is
one of the most significant (albeit difficult) tasks in a wide variety of physical scenarios.
For instance, the 
Bardeen-Cooper-Schrieffer ground state 
plays a pivotal r\^ole in condensed matter, 
being a condensate of Cooper pairs which
underpins the phenomenon of superconductivity~\cite{bcs}. 
In the same fashion, vacuum is 
crucially important to explain the spontaneous symmetry breaking~\cite{miransky} and the ensuing appearance of Nambu-Jona Lasinio~\cite{nambu}, Goldstone~\cite{goldstone,higgs} and pseudo-Goldstone bosons~\cite{pseudo}, both in low and high energy regimes. 
On the other hand, in QFT
vacuum energy is notoriously responsible for the existence of 
Casimir effect~\cite{casimir, Mitton}, which has been largely studied both
in flat~\cite{list1,list1b,list1c,list1d,list1e} and curved~\cite{list2b,list2c,list2d,list2e,list2f,list2g} spacetime in recent years. 
Moreover, the study of vacua in the presence of gravity 
leads to the loss of the absolute concept of particle and the emergence of distinctive phenomena such as the black hole radiation~\cite{hawking} and the akin Unruh effect~\cite{unruh}, which find application in many research areas~\cite{App}.

Starting from the above premises, in the present work we analyze some relevant and yet unexplored 
features of the flavor vacuum condensate, with a particular focus on its entanglement structure. 
For this purpose, we consider the case of mixing 
neutral bosonic fields. Besides the intrinsic importance (i.e. as in the case of meson mixing), this choice allows
to reduce several technical complications and render the physical insight
as transparent as possible. Given the mixing of two bosonic fields, in the following we will 
quantify the entanglement content
of the flavor vacuum by computing
von Neumann entanglement entropy of its reduced density matrix in the limit of small
mass difference and/or mixing angle.
We will then compare this result
with the entanglement 
of the thermal vacuum 
of Thermo Field Dynamics (TFD)~\cite{tfd,BTFD}, 
which exhibits the paradigmatic 
structure occurring in the case
of black holes~\cite{Israel} and
Unruh effect~\cite{unruh}, namely
the doubling of Fock space and 
the ensuing correlation between 
two different sets of modes 
(the particle states 
inside and outside the horizon
for Hawking-Unruh effect, the
physical and  auxiliary
modes in TFD).
It must be stressed that entanglement
in TFD has been previously discussed in Ref.~\cite{Hashizume} for 
both equilibrium and non-equilibrium states.
From the comparison between our results and the ones on TFD entanglement, 
we find that the condensate in the flavor vacuum 
is in general richer than the one in TFD, 
as it exhibits all types of contributions, both thermal and non-thermal. 
Finally, we determine a suitable limit in which
non-thermal contributions are subdominant, 
thereby allowing us to recognize a TFD-like structure in the flavor vacuum, 
with an effective temperature proportional to the mixing angle
and the mass difference between the two fields. 

The paper is organized as follows: in Sec.~\ref{BM} we review the main aspects of boson field mixing, focusing on the case of neutral scalar fields. 
In Sec.~\ref{EntFV} we quantify
the entanglement content of the flavor vacuum 
by computing the reduced von Neumann entropy,
and we discuss the condensate structure
of this state in connection with the thermal 
vacuum of TFD. Conclusions and outlook are provided in Sec.~\ref{Conc}. 
Throughout the work, we use natural units $c=\hbar=1$ 
while keeping the Boltzmann constant $k_B$ explicit. 

\section{Mixing of bosonic fields}
\label{BM}
In this Section, we review the crucial aspects associated with the
mixing of two scalar (spin-$0$) neutral fields. Clearly, the same considerations
can be extended to the case of three boson generations.
To this aim, we closely follow Refs.~\cite{bosonmix1,bosonmix2} and write down the mixing relations between fields with definite mass and flavor, that is\footnote{Strictly speaking, 
in the case of bosons we should refer
to the mixing of some other quantum number, 
such as the strangeness or the isospin rather
than the flavor. However, with an abuse of notation, in the following we
keep on denoting such intrinsic properties as ``flavor''  
and the corresponding mixed fields as ``flavor fields''. Furthermore, 
we work in a simplified two-flavor model.}
\bea
\label{mixing1}
\phi_A(x)&=&\cos\theta\,\phi_1(x)+\sin\theta\,\phi_2(x)\,,\\[2mm]\label{mixing}
\phi_B(x)&=&-\sin\theta\,\phi_1(x)+\cos\theta\,\phi_2(x)\,,
\eea
with an analogous set of equations for the conjugate momenta $\pi(x)=\pa_t\phi(x)$. The subscripts $A$ and $B$ indicate the fields in the flavor basis, whereas $1$ and $2$ the ones in the mass basis. Consequently, the expansions for $\phi_1$ and $\phi_2$ 
take the form
\be\label{fieldexp}
\phi_j(x)=\int\frac{d^3k}{\sqrt{2\lf(2\pi\ri)^3\omega_{k,j}}}\lf(a_{k,j}e^{-i\omega_{k,j}t}+a^\dagger_{-k,j}e^{i\omega_{k,j}t}\ri)e^{i\textbf{k}\cdot\textbf{x}}\,, \qquad j=1,2,
\ee
where $\omega_{k,j}=\sqrt{\textbf{k}^2+m_j^2}$
and $a_{k,j}$ $(a^\dagger_{k,j})$ are the bosonic
annihilation (creation) operators of field quanta
with momentum $k$ and mass $m_j$. 

By requiring that the fields and the conjugate
momenta obey the canonical commutation relation (CCR) at equal times, 
\be
\Bigl[\phi_i(x),\pi_j(x')\Bigr]_{t=t'}=i\hspace{0.2mm}\delta_{ij}\hspace{0.2mm}\delta\lf(\textbf{x}-\textbf{x}'\ri),
\ee
it follows that the only non-trivial commutator 
between the ladder operators is
\be
\left[a_{k,i}, a^\dagger_{k',j}\right]=\delta_{ij}\delta(\textbf{k}-\textbf{k}'). 
\ee
Let us now observe that Eqs.~(\ref{mixing1}) and (\ref{mixing}) and the ones for momenta can also be rewritten as
\bea
\phi_\si(x)&=&G_\theta^{-1}(t)\phi_j(x)G_\theta(t)\,,\\[2mm]\label{mixing2}
\pi_\si(x)&=&G_\theta^{-1}(t)\pi_j(x)G_\theta(t)\,,
\eea 
where $(\si,j)=\lf\{(A,1),(B,2)\ri\}$ and
\be\label{g}
G_\theta(t)=\mathrm{exp}\lf\{\theta\lf[S_+(t)-S_-(t)\ri]\ri\}
\ee
is the mixing generator and an element of $SU(2)$, whose algebra is built with the following operators
\bea\label{s+}
S_+(t)&=&-i\int d^3x\,\pi_1(x)\phi_2(x)\,,\\[2mm]\label{s-}
S_-(t)&=&-i\int d^3x\,\pi_2(x)\phi_1(x)\,\\[2mm]\label{s3}
S_3(t)&=&-\frac{i}{2}\int d^3x\lf[\pi_1(x)\phi_1(x)-\pi_2(x)\phi_2(x)\ri]\,.
\eea 
In light of the above equations, the flavor fields can then be expressed as
\be\label{flavfieldexp}
\phi_\si(x)=\int\frac{d^3k}{\sqrt{2\lf(2\pi\ri)^3\,\omega_{k,j}}}\lf(a_{k,\si}(t)\hspace{0.2mm}e^{-i\omega_{k,j}t}+a^\dagger_{-k,\si}(t)\hspace{0.2mm}e^{i\omega_{k,j}t}\ri)e^{i\textbf{k}\cdot\textbf{x}}\,,\qquad \sigma=A,B,
\ee
and, with the aid of Eqs.~(\ref{mixing1}) and (\ref{mixing}), we recognize the Bogoliubov transformation between the flavor and mass ladder operators
\bea
\label{bog}
a_{k,A}(t)&=&\cos\theta\,a_{k,1}+\sin\theta\lf[U_k^*(t)\,a_{k,2}+V_k(t)\,a^\dagger_{-k,2}\ri],\\[2mm]
a_{k,B}(t)&=&\cos\theta\,a_{k,2}-\sin\theta\lf[U_k(t)\,a_{k,1}-V_k(t)\,a^\dagger_{-k,1}\ri].
\label{bog2}
\eea
The above relations exhibit the structure of rotations
nested into Bogoliubov transformations with coefficients
$U_k(t)$ and $V_k(t)$ given by
\be\label{uv}
U_k(t)=\lf|U_k\ri|e^{i\lf(\omega_{k,2}-\omega_{k,1}\ri)t}\,, \quad V_k(t)=\lf|V_k\ri|e^{i\lf(\omega_{k,1}+\omega_{k,2}\ri)t},
\ee
\be\label{uvmod}
\lf|U_k\ri|=\frac{1}{2}\lf(\sqrt{\frac{\omega_{k,1}}{\omega_{k,2}}}+\sqrt{\frac{\omega_{k,2}}{\omega_{k,1}}}\ri)\,, \quad \lf|V_k\ri|=\frac{1}{2}\lf(\sqrt{\frac{\omega_{k,1}}{\omega_{k,2}}}-\sqrt{\frac{\omega_{k,2}}{\omega_{k,1}}}\ri)\,.
\ee 
Accordingly, 
the flavor vacuum is provided with 
with a $SU(2)$ coherent state structure~\cite{perelomov}
\be\label{flavac}
|00(t)\rangle_{A,B}=G_\theta^{-1}(t)|00\rangle_{1,2}\,, 
\ee
with a condensation density given by
\begin{equation}
\label{eqn:denscondbosoni2} 
{}_{A,B}\langle 00(t)| a_{k,j}^{\dagger}\, a_{k,j} |00(t)\rangle_{A,B}\,=\, \sin^{2}\theta \lf|V_k\ri|^2, \qquad j=1,2\,.
\end{equation}
In the infinite volume limit~\cite{bosonmix1,bosonmix2}, the two sets of vacua become orthogonal, namely ${}_{1,2}\langle00|00(t)\rangle_{A,B}\overset{V\to\infty}{\longrightarrow}0$, $\forall t$, giving rise to physically inequivalent Fock spaces (i.e. unitarily inequivalent representations of the canonical commutation relations for fields). 

Let us now cast the mixing generator in terms of mass definite annihilators and creators. Straightforward calculations lead to
\be\label{momflavacfirst}
|00(t)\rangle_{A,B}=\exp\Bigl\{-\theta\int d^3k\lf[U^*_k(t)\,a^\dagger_{k,1}a_{k,2}-U_k(t)\,a_{k,1}a^\dagger_{k,2}+V_k(t)\,a^\dagger_{k,1}a^\dagger_{-k,2}-V_k^*(t)\,a_{k,1}a_{-k,2}\ri]\Bigr\}|00\rangle_{1,2}\,.
\ee
Without harming the generality of our results, henceforth 
we perform calculations for $t=0$. Thus, we have $|00(t=0)\rangle_{A,B}\equiv|00\rangle_{A,B}$ with\footnote{Notice that, for $t=0$, $U_k(t)=\lf|U_k\ri|$ and $V_k(t)=\lf|V_k\ri|$ (see Eq.~\eqref{uv}).}
\be\label{momflavac}
|00\rangle_{A,B}=\exp\Bigl\{-\theta\int d^3k\lf[U_k\lf(a^\dagger_{k,1}a_{k,2}-a_{k,1}a^\dagger_{k,2}\ri)+V_k\lf(a^\dagger_{k,1}a^\dagger_{-k,2}-a_{k,1}a_{-k,2}\ri)\ri]\Bigr\}|00\rangle_{1,2}\,.
\ee
Equation~(\ref{momflavac}) allows us to immediately identify the generator of the rotation (the operator in the brackets which multiplies $U_k$) and the one responsible for the Bogoliubov transformation (the operator in the brackets which multiplies $V_k$), in complete agreement with the fermion case~\cite{bogoliubov}.

For later convenience, we can now further manipulate
Eq.~\eqref{momflavac} by considering the case of a discrete set
of modes. Apart from an irrelevant numerical factor, Eq.~(\ref{momflavac}) becomes
\be\label{finitevac}
|00\rangle_{A,B}=\exp\Bigl\{-\theta\sum_k\lf[U_k\lf(a^\dagger_{k}b_{k}-a_{k}b^\dagger_{k}\ri)+V_k\lf(a^\dagger_{k}b^\dagger_{-k}-a_{k}b_{-k}\ri)\ri]\Bigr\}|00\rangle_{1,2}\,.
\ee
where we have introduced the
shorthand notation $a_{k,1}\equiv a_k$ and $a_{k,2}\equiv b_k$.
In addition, since the transformations Eqs.~(\ref{mixing1}) and (\ref{mixing}) are valid  for any value of the rotation angle, 
we can focus on the case of small values of $\theta$ to the $\mathcal{O}(\theta^2)$-order, at which the first non-trivial contribution is expected. 
In this framework, we can use Zassenhaus formula
\be\label{zassenhaus}
e^{\lambda\lf(X+Y\ri)}=e^{\lambda X}e^{\lambda Y}e^{-\frac{\lambda^2}{2}\lf[X,Y\ri]}\,,
\ee
up to $\mathcal{O}(\lambda^2)$ to approximate the flavor vacuum in Eq.~\eqref{finitevac}. In fact, if we identify $\lambda\to-\theta$, $X\to\sum_kU_k\lf(a^\dagger_kb_k-a_kb^\dagger_k\ri)$ and $Y\to\sum_kV_k\lf(a^\dagger_kb^\dagger_{-k}-a_kb_{-k}\ri)$, we are led to
\bea
\label{finitevac2}
|00\rangle_{A,B}&=&e^{-\theta\sum_kU_k\lf(a^\dagger_kb_k-a_kb^\dagger_k\ri)}e^{-\theta\sum_kV_k\lf(a^\dagger_kb^\dagger_{-k}-a_kb_{-k}\ri)}e^{-\frac{\theta^2}{2}\sum_kU_kV_k\lf(a^\dagger_ka^\dagger_{-k}-b^\dagger_kb^\dagger_{-k}-a_ka_{-k}+b_kb_{-k}\ri)}|00\rangle_{1,2}\\[2mm]\non
&=&e^{-\frac{\theta^2}{2}\sum_kU_kV_k\lf(a^\dagger_ka^\dagger_{-k}-b^\dagger_kb^\dagger_{-k}\ri)}e^{-\theta\sum_kU_k\lf(a^\dagger_kb_k-a_kb^\dagger_k\ri)}e^{-\theta\sum_kV_k\lf(a^\dagger_kb^\dagger_{-k}-a_kb_{-k}\ri)}|00\rangle_{1,2}\,,
\eea
where in the second step we have made use of a simplification allowed only in the given approximation for $\theta$. 
Furthermore, by applying the identity~(\ref{zassenhaus}) to the two operators in Eq.~(\ref{finitevac2}) that depend on $U_k$ and $V_k$, we get 
\bea
\label{newforvac}
|00\rangle_{A,B}&=&e^{-\frac{\theta^2}{2}\sum_kU_kV_k\lf(a^\dagger_ka^\dagger_{-k}-b^\dagger_kb^\dagger_{-k}\ri)}e^{-\theta\sum_kU_ka^\dagger_kb_k}e^{\theta\sum_kU_ka_kb^\dagger_k}e^{\frac{\theta^2}{2}\sum_kU_k^2\lf(a_ka_k^\dagger-b_kb_k^\dagger\ri)}\\[2mm]\non
&\times&e^{-\theta\sum_kV_ka^\dagger_kb^\dagger_{-k}}e^{\theta\sum_kV_ka_kb_{-k}}e^{-\frac{\theta^2}{2}\sum_kV_k^2\lf(a_ka_k^\dagger+b_k^\dagger b_k\ri)}|00\rangle_{1,2}\\[2mm]\label{finitevac3}\non
&\simeq&e^{-\frac{\theta^2}{2}\sum_kU_kV_k\lf(a^\dagger_ka^\dagger_{-k}-b^\dagger_kb^\dagger_{-k}\ri)}e^{-\frac{\theta^2}{2}\sum_k\lf(V_k^2b^\dagger_kb_k+U_k^2b_kb^\dagger_k\ri)}e^{-\theta\sum_kV_ka^\dagger_kb^\dagger_{-k}}|00\rangle_{1,2}\,.
\eea
In performing the second passage, we have omitted an unimportant constant factor and we have made use of the current approximation to streamline the shape of the total operator. 
We will take advantage of the form~\eqref{newforvac} of the flavor vacuum
in the next Section, when we compare the
condensate structure of this state to
the one of the thermal vacuum of Thermo Field Dynamics.

\section{Entanglement of the flavor vacuum}
\label{EntFV}
Let us now quantify the entanglement between the massive particle states  
in the flavor vacuum. As said earlier, we consider the case $t=0$. 
For notational convenience,
it comes in handy to rewrite Eq.~\eqref{momflavac} 
as
\be\label{state}
|00\rangle_{A,B}\equiv|\psi\rangle=\exp\Bigl\{-\theta\int d^3k\lf[U_k\lf(a^\dagger_{k}b_{k}-a_{k}b^\dagger_{k}\ri)+V_k\lf(a^\dagger_{k}b^\dagger_{-k}-a_{k}b_{-k}\ri)\ri]\Bigr\}|00\rangle\,, 
\ee
where
$|00\rangle\equiv|00\rangle_{1,2}$
and we have used the same notation
as in Eq.~\eqref{finitevac} for ladder operators. 
Interestingly, we observe that the above transformation is the result of a simultaneous coexistence of a beam splitter and a two-mode squeezing transformation.

As a preliminary analysis, we can perform a first-order approximation in $\theta$ and in the case of small mass difference, namely when $m_1=m$, $m_2=m+\de m$ and hence $\varepsilon\equiv(m_2-m_1)/m_1=\de m/m\ll 1$. This is in line with many works regarding flavor mixing, and in particular with Ref.~\cite{bogoliubov}, in which the neutrino flavor vacuum is written up to the second order in $\theta$ and $\varepsilon$. Here, we make the same considerations with the purpose of seeking an analogous result.

Starting from Eq.~(\ref{state}), it is immediate to derive that
\be\label{state2}
|\psi\rangle=\exp\{-\theta\int d^3k\lf[\lf(a_k^\dagger b_k-a_kb_k^\dagger\ri)-\frac{\varepsilon}{2}\frac{m^2}{\omega_k^2}\lf(a_k^\dagger b_{-k}^\dagger-a_kb_{-k}\ri)\ri]\}|00\rangle\,,
\ee
and thus expand the exponential operator as
\be\label{state3}
|\psi\rangle=\sum_{n=0}^\infty\frac{1}{n!}\lf\{-\theta\int d^3k\lf[\lf(a_k^\dagger b_k-a_kb_k^\dagger\ri)-\frac{\varepsilon}{2}\frac{m^2}{\omega_k^2}\lf(a_k^\dagger b_{-k}^\dagger-a_kb_{-k}\ri)\ri]\ri\}^n|00\rangle\,.
\ee 
Since terms of the order $\mathcal{O}\lf(\varepsilon^2\ri)$ are neglected, we observe that it is possible to identify two recursive formulas in the expression~(\ref{state3}), one in front of the operator $a^\dagger b^\dagger$ and another one next to $\lf(a^\dagger a^\dagger-b^\dagger b^\dagger\ri)$. The exact computation yields
\be\label{state4}
|\psi\rangle=\lf\{\textbf{1}+\frac{\varepsilon}{2}\lf[\sum_{n=0}^\infty\frac{(-1)^{n+1}\theta^{2n+2}\,4^n}{(2n+2)!}\ri]\int d^3k\,\frac{m^2}{\omega_k^2}\lf[a^\dagger_ka^\dagger_{-k}-b^\dagger_kb^\dagger_{-k}\ri]+\frac{\varepsilon}{2}\lf[\sum_{n=0}^\infty\frac{(-1)^n\theta^{2n+1}\,4^n}{(2n+1)!}\ri]\int d^3k\,\frac{m^2}{\omega_k^2}\,a^\dagger_kb^\dagger_{-k}\ri\}|00\rangle\,.
\ee
It is straightforward to realize that the series in Eq.~(\ref{state4}) converge to two simple analytic functions:
\begin{eqnarray}
\sum_{n=0}^\infty\frac{(-1)^{n+1}\theta^{2n+2}\,4^n}{(2n+2)!}&=&-\frac{\sin^2\theta}{2}\,,\\[2mm]
\sum_{n=0}^\infty\frac{(-1)^n\theta^{2n+1}\,4^n}{(2n+1)!}&=&\frac{\sin2\theta}{2}\,.
\end{eqnarray}
As a result, in the limit of small mass difference the flavor vacuum  takes the form
\be\label{momentafin}
|\psi\rangle=\lf\{\textbf{1}-\frac{\varepsilon}{4}\sin^2{\theta}\int d^3k\,\frac{m^2}{\omega_k^2}\lf[a^\dagger_ka^\dagger_{-k}-b^\dagger_kb^\dagger_{-k}\ri]+\frac{\varepsilon}{4}{\sin{2}\theta}\int d^3k\,\frac{m^2}{\omega_k^2}\,a^\dagger_kb^\dagger_{-k}\ri\}|00\rangle\,.
\ee
Before proceeding with the computation of von Neumann
entropy, we pause to compare the above
form of the flavor vacuum
with the thermal vacuum of Thermo Field Dynamics 
defined in the Appendix (see in particular Eq.~\eqref{thetavac2}). 
The similarity between these two states
sinks its roots in the fact that in TFD there is the need to double the physical Hilbert space by introducing a dual space (and hence an auxiliary field) whose excitations are holes from the point of view of the physical field (a detailed mathematical explanation of these notions can be found in the Appendix).
As a consequence, the thermal vacuum appears as
a condensate of excitations of the physical and 
auxiliary fields, which thus resembles the structure of the flavor
vacuum being a condensate of particle/antiparticle
pairs with different masses.

In this vein, we emphasize
that a first attempt to perform a comparison 
between the flavor and thermal vacua
was carried out in Ref.~\cite{Dimauro} and later
in~\cite{bogoliubov}, arguing that the two states
cannot be exactly matched, since the would-be entropy operator 
defined for mixed fields does not possess the same properties as the one introduced in TFD (see Eq.~(\ref{ent})). 
However, here we tackle this problem from a different perspective;
indeed, we directly look at the inherent structure of the two vacua.
Even though the result of Ref.~\cite{bogoliubov} remains in general valid
since the thermal state~\eqref{thetavac2}
does not contain terms of the form $a_k^\dagger a_{-k}^\dagger$, $b_k^\dagger b_{-k}^\dagger$ which instead appear in the
flavor vacuum~\eqref{momentafin}, for small values of the mixing
angle we can approximate Eq.~\eqref{momentafin} to the leading order as
\be
\label{abst}
|\psi\rangle=\lf\{\textbf{1}+\frac{\varepsilon}{2}{\theta}\int d^3k\,\frac{m^2}{\omega_k^2}\,a^\dagger_kb^\dagger_{-k}\ri\}|00\rangle.
\ee
Then, from comparison with the 
TFD vacuum in Eq.~\eqref{neweq}, 
\be
\non
|0(\vartheta)\rangle\simeq\textbf{1}+\sum_ka^\dagger_k\tilde a^\dagger_k\vartheta_k|0\rangle\,,
\ee
one can recognize in the state~\eqref{abst} a thermal-like vacuum
with an inverse temperature $\beta=(k_BT)^{-1}$ given by (up to 
a scaling factor due to the conversion of the $k$-integral
into a discrete sum, as viewed in Eq.~\eqref{finitevac} and following)
\be
\label{Tmix}
\beta\simeq\frac{1}{\omega_k}\ln\left(\frac{2\hspace{0.2mm}\omega_k^2}{\varepsilon\hspace{0.2mm}\theta\hspace{0.2mm}m^2}\right).
\ee
Clearly, for $\theta$ and/or $\varepsilon=0$, 
we have $\beta\rightarrow\infty$, or
equivalently $T\rightarrow0$. This is somehow expected
since, for vanishing mixing, the flavor and mass vacua
coincide with each other,  
which corresponds in the TFD language
to the case where the doubling of degrees of freedom (and thus the
temperature) disappears.

Let us now come back to
the quantification of  von Neumann entropy
of the flavor vacuum. To this aim,  we focus
on the $k=0$ mode, for which 
the condensation density 
reaches its maximal value
in the case of boson mixing (see Eq.~\eqref{eqn:denscondbosoni2})~\cite{bosonmix2}. 
Therefore, we expect that the effects
of the mixing transformation are
maximally non-trivial in this case.
However, for this purpose  
we need to go beyond the linear order approximation
in the mass difference, as it can be easily shown that the
linear term in $\varepsilon$ gives a vanishing contribution. 
Then, by resorting to Eq.~(\ref{state3}) and retaining only the factors that do not exceed $\mathcal{O}(\varepsilon^2)$,  we find a recurrence series of the form 
\bea
\label{finstate2}
&&\hspace{-4mm}|\psi\rangle=\Bigl\{\textbf{1}+\lf(\frac{\varepsilon}{2}-\frac{\varepsilon^2}{4}\ri)\lf[\sum_{n=0}^\infty\frac{(-1)^n\theta^{2n+1}\,4^n}{(2n+1)!}\ri]a^\dagger b^\dagger+\lf(\frac{\varepsilon}{2}-\frac{\varepsilon^2}{4}\ri)\lf[\sum_{n=0}^\infty\frac{(-1)^{n+1}\theta^{2n+2}\,4^n}{(2n+2)!}\ri]\lf[(a^\dagger)^2-(b^\dagger)^2\ri]\\[2mm]\non
&&\hspace{-0.4cm}+\,\frac{\varepsilon^2}{4}\lf[\sum_{n=0}^\infty\frac{(-1)^{n}\theta^{2n+2}\,4^n}{(2n+2)!}\ri]\lf[(a^\dagger)^2(b^\dagger)^2-\textbf{1}\ri]+\frac{3\,\varepsilon^2}{4}\lf\{\sum_{n=0}^\infty\lf[\frac{(-1)^{n+1}\theta^{2n+3}4^n}{(2n+3)!}\sum_{m=0}^n\lf(3\sqrt{2}+\sqrt{6}\ri)^m\ri]\ri\}\lf[(a^\dagger)^3b^\dagger-a^\dagger(b^\dagger)^3\ri]\\[2mm]\non
&&\hspace{-0.4cm}+\,\frac{3\,\varepsilon^2}{4}\lf\{\sum_{n=0}^\infty\lf[\frac{(-1)^{n}\theta^{2n+4}4^n}{(2n+4)!}\sum_{m=0}^n\lf(3\sqrt{2}+\sqrt{6}\ri)^m\ri]\ri\}\lf[(a^\dagger)^4-6\sqrt{2}(a^\dagger)^2(b^\dagger)^2+(b^\dagger)^4\ri]\Bigr\}|00\rangle\,.
\eea
The series appearing in Eq.~(\ref{finstate2}) converge to trigonometric functions in $\theta$. More precisely,  we have 
\bea\label{finstate3}
&|\psi\rangle&=\lf[1-\frac{\varepsilon^2}{8}\sin^2{\theta}\ri]|00\rangle-\lf(\frac{\varepsilon}{4}-\frac{\varepsilon^2}{8}\ri)\sin^2{\theta}\Bigl(|20\rangle-|02\rangle\Bigr)+\lf(\frac{\varepsilon}{4}-\frac{\varepsilon^2}{8}\ri)\sin{2}\theta\,|11\rangle\\[2mm]\non
&+&\frac{\varepsilon^2}{4}\lf[\frac{\sin^2\theta}{2}-18\sqrt{2}\,\chi(\theta)\ri]|22\rangle+\frac{3\,\varepsilon^2}{4}\chi(\theta)\Bigl(|40\rangle+|04\rangle\Bigr)+\frac{3\,\varepsilon^2}{4}\xi(\theta)\Bigl(|31\rangle-|13\rangle\Bigr)\,,
\eea
where
\bea\label{chi}
\chi(\theta)&=&\frac{\lf(6+2\sqrt{3}\ri)\lf(1-\cos2\theta\ri)-\sqrt{2}\lf[1-\cos\lf(2^{\frac{5}{4}}\sqrt{3+\sqrt{3}}\,\theta\ri)\ri]}{32\lf(3+\sqrt{3}\ri)\lf(3\sqrt{2}+\sqrt{6}-1\ri)}\,,\\[3mm]\label{xi}
\xi(\theta)&=&\frac{2^{\frac{3}{4}}\sin\lf(2^{\frac{5}{4}}\sqrt{3+\sqrt{3}}\,\theta\ri)-2\sqrt{3+\sqrt{3}}\sin2\theta}{16\sqrt{3+\sqrt{3}}\lf(3\sqrt{2}+\sqrt{6}-1\ri)}\,.
\eea
The state in Eq.~(\ref{finstate3}) is not normalized, since $\langle\psi|\psi\rangle\neq1$. For this reason, before computing the projector $\rho=|\psi\rangle\langle\psi|$, we must divide the above state by a suitable normalization factor, which is explicitly given by
\be\label{N}
N=\sqrt{\langle\psi|\psi\rangle}=\sqrt{1-\frac{\varepsilon^2}{4}\lf[\sin^2{\theta}-\frac{\sin^4{\theta}}{2}-\frac{\sin^2{2}\theta}{4}\ri]}\,.
\ee
Accordingly, the normalized density matrix reads 
\bea
\label{ndensity}
&\rho&=\frac{|\psi\rangle\langle\psi|}{N^2}=\lf[1-\frac{\varepsilon^2}{8}\sin^4{\theta}-\frac{\varepsilon^2}{16}\sin^2{2}\theta\ri]|00\rangle\langle00|+\lf(\frac{\varepsilon}{4}-\frac{\varepsilon^2}{8}\ri)\sin{2}\theta\Bigl(|00\rangle\langle11|+|11\rangle\langle00|\Bigr)\\[2mm]\non
&-&\lf(\frac{\varepsilon}{4}-\frac{\varepsilon^2}{8}\ri)\sin^2{\theta}\Bigl(|00\rangle\langle20|-|00\rangle\langle02|+|20\rangle\langle00|-|02\rangle\langle00|\Bigr)+\frac{\varepsilon^2}{16}\sin^2{2}\theta\,|11\rangle\langle11|\\[2mm]\non
&+&\frac{\varepsilon^2}{4}\lf[\frac{\sin^2\theta}{2}-18\sqrt{2}\,\chi(\theta)\ri]\Bigl(|00\rangle\langle22|+|22\rangle\langle00|\Bigr)+\frac{3\,\varepsilon^2}{4}\chi(\theta)\Bigl(|00\rangle\langle40|+|00\rangle\langle04|+|40\rangle\langle00|+|04\rangle\langle00|\Bigr)\\[2mm]\non
&+&\frac{\varepsilon^2}{16}\sin^4{\theta}\Bigl(|20\rangle\langle20|-|20\rangle\langle02|-|02\rangle\langle20|+|02\rangle\langle02|\Bigr)+\frac{3\,\varepsilon^2}{4}\xi(\theta)\Bigl(|00\rangle\langle31|-|00\rangle\langle13|+|31\rangle\langle00|-|13\rangle\langle00|\Bigr)\\[2mm]\non
&-&\frac{\varepsilon^2}{16}\sin^2{\theta}\sin{2}\theta\Bigl(|20\rangle\langle11|-|02\rangle\langle11|+|11\rangle\langle20|-|11\rangle\langle02|\Bigr)\,.
\eea
By partial tracing $\rho$ with respect to either one of the two subsystems, say $\mathcal{S}_1$, we obtain the reduced density matrix 
\bea
\label{nreduc}
\rho^{(2)}_r=\mathrm{Tr}_{\mathcal{S}_1}\rho&=&\lf[1-\frac{\varepsilon^2}{16}\sin^22\theta-\frac{\varepsilon^2}{16}\sin^4\theta\ri]|0\rangle\langle0|+\lf(\frac{\varepsilon}{4}-\frac{\varepsilon^2}{8}\ri)\sin^2\theta\Bigl(|0\rangle\langle2|+|2\rangle\langle0|\Bigr)+\frac{\varepsilon^2}{16}\sin^22\theta\,|1\rangle\langle1|\\[2mm]\non
&+&\frac{3\,\varepsilon^2}{4}\chi(\theta)\Bigl(|0\rangle\langle4|+|4\rangle\langle0|\Bigr)+\frac{\varepsilon^2}{16}\sin^4\theta\,|2\rangle\langle2|\,,
\eea
with the same result for $\rho^{(1)}_r=\mathrm{Tr}_{\mathcal{S}_2}\rho$.
From Eq.~\eqref{nreduc}, we can compute von Neumann entropy up to $\mathcal{O}(\varepsilon^2)$. Notice that the matrix $\rho^{(1)}_r=\rho^{(2)}_r\equiv\rho_r$, given by
\be\label{nreduc2}\renewcommand{\arraystretch}{1.5}
\rho_r=\begin{pmatrix} 1-\frac{\varepsilon^2}{16}\sin^22\theta-\frac{\varepsilon^2}{16}\sin^4\theta & 0 & \lf(\frac{\varepsilon}{4}-\frac{\varepsilon^2}{8}\ri)\sin^2\theta & \frac{3\,\varepsilon^2}{4}\chi(\theta) \\ 0 & \frac{\varepsilon^2}{16}\sin^22\theta & 0 & 0 \\ \lf(\frac{\varepsilon}{4}-\frac{\varepsilon^2}{8}\ri)\sin^2\theta & 0 & \frac{\varepsilon^2}{16}\sin^4\theta & 0 \\ \frac{3\,\varepsilon^2}{4}\chi(\theta) & 0 & 0 & 0 \end{pmatrix},
\ee
has eigenvalues
\be\label{eigennew}
\lambda_i=\lf\{\mathcal{O}(\varepsilon^3), \mathcal{O}(\varepsilon^3), 1-\frac{\varepsilon^2}{32}\lf(1-\cos4\theta\ri), \frac{\varepsilon^2}{16}\sin^22\theta\ri\}\,.
\ee
Therefore, to order $\varepsilon^2$, von Neumann entanglement entropy reads
\be
\label{final}
S_V= -\sum_i \lambda_i \log_2{\lambda_i}=-\lf[1-\frac{\varepsilon^2}{32}\lf(1-\cos4\theta\ri)\ri]\log_2\lf[1-\frac{\varepsilon^2}{32}\lf(1-\cos4\theta\ri)\ri]-\frac{\varepsilon^2}{16}\sin^22\theta\log_2\lf(\frac{\varepsilon^2}{16}\sin^22\theta\ri)\,.
\ee
\begin{figure}[t]
\centering
\includegraphics[width=14cm]{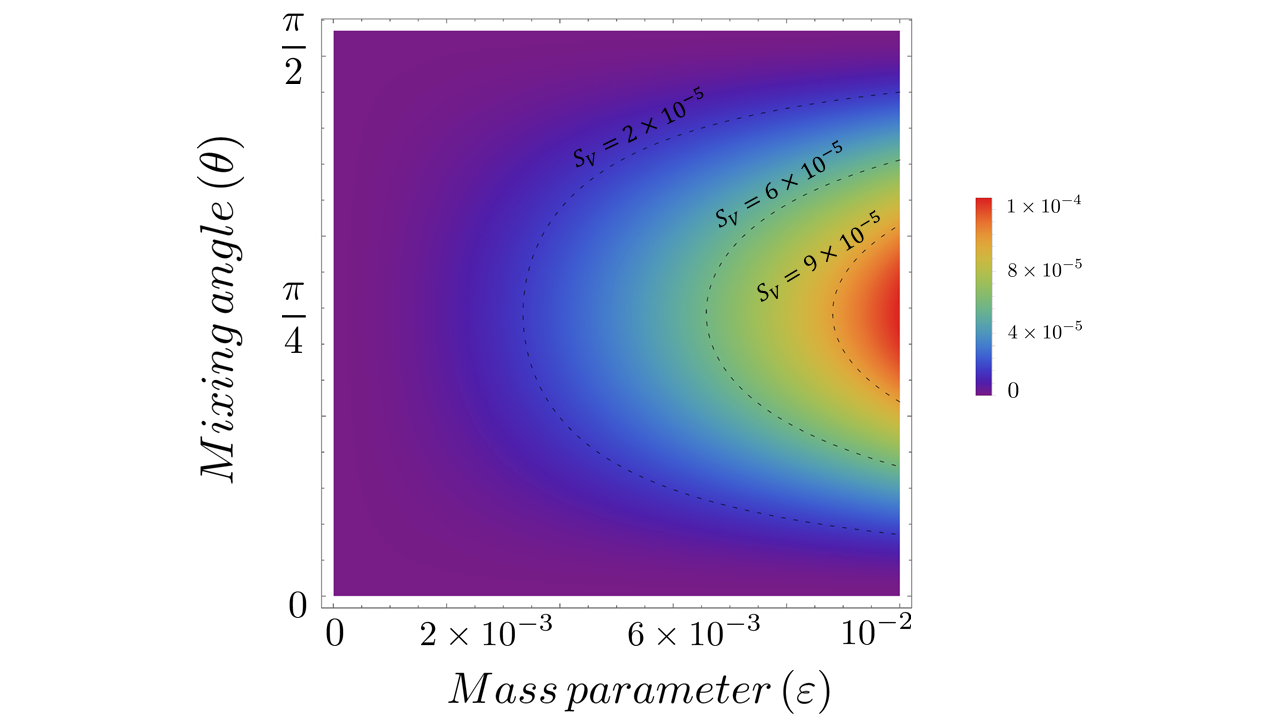}
\caption{Behavior of von Neumann entanglement entropy $S_V$ as a function of $\varepsilon\in\lf[0, 10^{-2}\ri]$ and $\theta\in\lf[0, \frac{\pi}{2}\ri]$. As explicitly shown by the level curves, $S_V$ monotonically decreases with decreasing $\varepsilon$.} 
\label{figure}
\end{figure}
As expected, in the limit of either $\varepsilon\to0$ or $\theta\to0$, we recover $S_V=0$, as it should correctly be, because in these cases the flavor vacuum reduces to the tensor product $|00\rangle_{1,2}=|0\rangle_1\otimes|0\rangle_2$. In Fig.~\ref{figure} we report the behavior of von Neumann entropy as a function of both $\varepsilon$ and $\theta$. In order to better convey the dependence of $S_V$ on the mixing angle, it is also appropriate to exhibit how it varies by keeping $\varepsilon$ fixed. This is done in Fig.~\ref{figure2}. In particular, we can notice that for arbitrary $\varepsilon$ the maximum of $S_V$ always occurs at the perfectly balanced mixing angle $\theta_{max}=\pi/4$. Increasing $\varepsilon$ only results in an upper shift in the magnitude of $S_V$, with an essentially invariant shape. The value of $\theta_{max}$ is exactly the one that could have been initially predicted, since $\pi/4$ corresponds to maximal mixing.

\section{Conclusions}
\label{Conc}
In this paper, we have explored some non-trivial features of the flavor vacuum in the case of mixing of two neutral scalar fields. A first preliminary analysis has shown that we cannot quantify entanglement by resorting to a direct comparison between $|00\rangle_{A,B}$ and the free thermal vacuum introduced in Thermo Field Dynamics, since the nature of the two states is not exactly the same. Note that this is consistent with the result found in Ref.~\cite{bogoliubov} in the context of neutrino mixing. 
A similar outcome has also been exhibited in Refs.~\cite{prd}, 
where it has been shown that the flavor vacuum
for an accelerated observer is not strictly a thermal state, 
in contrast with the case of the standard Unruh effect.
However, in spite of these differences, we 
have found that, in a proper limit, the flavor vacuum
can be identified with the vacuum of TFD, with a temperature
dependent on the mixing angle and the 
mass difference of the two mixed fields. 

\begin{figure}[t]
\centering
\includegraphics[width=15cm]{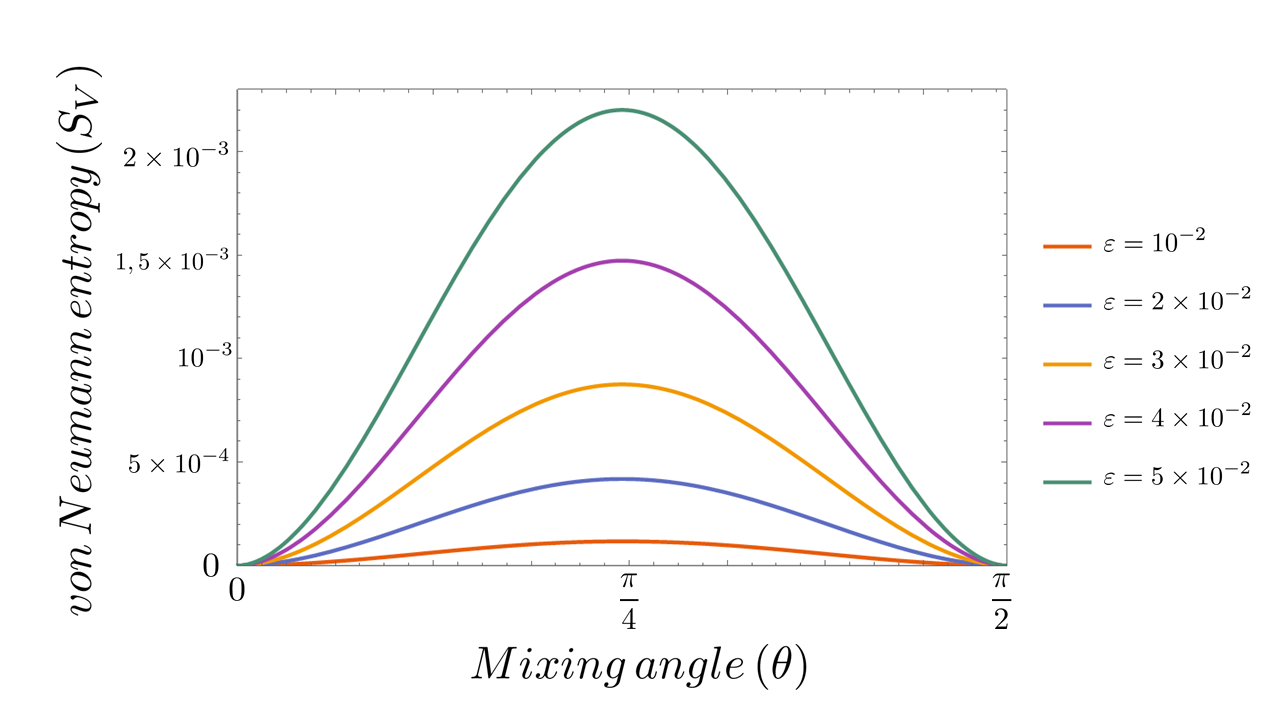}
\caption{Behavior of von Neumann entanglement entropy $S_V$ as a function of $\theta\in\lf[0, \frac{\pi}{2}\ri]$ for different values of $\varepsilon$.}
\label{figure2}
\end{figure}

To further investigate the condensate structure of the flavor vacuum, 
we have restricted our attention to the mode $k=0$, for 
which the condensation density in $|00\rangle_{A,B}$ is maximal. In this setting,  we have quantified von Neumann entropy for small values of the mass difference. 
The shape exhibited in Fig.~\ref{figure} is the one derived from our analysis, and the picture of Fig.~\ref{figure2} shows that there is an angle in correspondence of which von Neumann entropy is maximal for a given value of the small mass difference. It is worth observing that such angle is precisely the one responsible for maximal mixing, namely $\theta_{max}=\pi/4$. Clearly, it would be interesting
to go beyond the single-mode analysis and derive the full expression for the
von Neumann entropy. However, we expect that the 
result will exhibit some ultraviolet divergence 
of the same kind as in Ref.~\cite{Capozz}, 
where the contribution of the flavor vacuum energy to
the cosmological constant was shown to diverge only logarithmically, 
in contrast with the standard asymptotic behavior of
free-field vacuum energy. 

It is worth remarking that entanglement entropy and particle mixing have already been analyzed together in several papers in recent years~\cite{entmix,entmix2}, but only in the context of flavor transitions and for one-particle flavor states. In this respect, it is interesting to compare the field-theoretical approach to flavor entanglement developed in the present work with the exact methods for the quantification of entanglement introduced in the framework of quantum information and quantum optics of non-relativistic continuous variable systems (see for instance Refs.~\cite{gaussian,gaussian2,gaussian3,review} and references therein). Such methods revolve around the transformation of the quadrature operators (namely, position and momentum) under a given operation, which in our case is the mixing transformation. Since at the level of these operators mixing acts merely as a rotation, we would not achieve the desired result, thus preventing us from reaching an accurate evaluation of the entanglement entropy of the flavor vacuum via the procedures adopted for quantum optical systems~\cite{gaussian,gaussian2,gaussian3}.

Finally, we emphasize that the present analysis may have several implications
in a wide range of diversified contexts. For instance, once fully
extended to non-inertial frames~\cite{accCa,accAl,acc1} 
and, more generally, to gravity scenarios~\cite{gravity,gravity2,gravity3,gravity4,gravity5,gravity6,gravity7,gravity8,gravity9,gravity10,gravity11}, 
one could investigate how the information content of the flavor vacuum is degraded for increasing values of the acceleration/gravity. 
Note that a similar analysis 
has been carried out in Ref.~\cite{fuentes1} 
for the case of the entanglement between two free
modes of a scalar field as seen by an inertial observer
detecting one of the modes and a uniformly accelerated
observer detecting the other one. These and other aspects
will be thoroughly investigated in future works. 

\appendix
\section{Thermo Field Dynamics}
\label{TFD}
This Appendix is devoted to review the basics of 
Thermo-Field-Dynamics (TFD)~\cite{tfd}, which represents
one of the approaches to QFT at finite temperature
and density. The reason for this study
lies in the fact the TFD and the 
formalism of mixing analyzed in Sec.~\ref{BM} 
exhibit some non-trivial conceptual similarities, 
the most prominent one being the doubling
of the degrees of freedom of the original system. 

In order to introduce the TFD from a more quantitative point of view, 
let us consider the Hamiltonian of a free physical system written in the usual form
\be\label{freeham}
H=\sum_k\epsilon_ka^\dagger_ka_k\,.
\ee 
We now define a completely identical fictitious system, which we name ``tilde'' system, sharing the same properties of the physical one. The former can then be written as
\be\label{tildeham}
\tilde{H}=\sum_k\epsilon_k\tilde{a}^\dagger_k\tilde{a}_k\,.
\ee
The above relation allows us to establish
a new set of ``thermal'' operators via  the Bogoliubov transformations
\bea
\label{tildebog12}
a_k(\vartheta)&=&a_k\cosh\vartheta_k-\tilde{a}^\dagger_k\sinh\vartheta_k\,,\\[2mm]
\label{tildebog22}
\tilde{a}_k(\vartheta)&=&\tilde{a}_k\cosh\vartheta_k-a^\dagger_k\sinh\vartheta_k\,,
\eea
with $\theta_{k}$ encoding the information
about the original and thermal ladder operators.
Notice that the transformations~\eqref{tildebog12} and~\eqref{tildebog22}
leave the total Hamiltonian of the system $H_{tot}=H-\tilde{H}$ invariant, 
provided that the Hamiltonians~(\ref{freeham}) and~(\ref{tildeham}) 
have the same spectrum.

The Bogoliubov transformation~\eqref{tildebog12} and~\eqref{tildebog22} can also be recast as
\bea
a_k(\vartheta)&=&e^{-iG}\,a_k\,e^{iG}\,,\\[2mm]\label{tildebog2}
\tilde{a}_k(\vartheta)&=&e^{-iG}\,\tilde{a}_k\,e^{iG}\,,
\eea
with
\be\label{generatore}
G=i\sum_k\vartheta_k\lf(a_k^\dagger\tilde{a}^\dagger_k-\tilde{a}_ka_k\ri)\,,
\ee
being the generator of the transformation, which is also a conserved quantity (namely $\lf[G,H_{tot}\ri]=0$) since the total Hamiltonian is conserved under its action. By means of $G$, we can now build a new vacuum state for the operators $a_k(\vartheta)$ and $\tilde{a}_k(\vartheta)$ defined as~\cite{tfd}
\be\label{thetavac}
|0(\vartheta)\rangle=e^{-iG}|0\rangle\,,
\ee
where $|0\rangle$ is the vacuum associated with $a_k$ and $\tilde{a}_k$. If we explicitly act with the above operator on $|0\rangle$, we are left with
\be\label{thetavac2}
|0(\vartheta)\rangle=\prod_k\frac{e^{a^\dagger_k\tilde{a}^\dagger_k\tanh\vartheta_k}}{\cosh\vartheta_k}\,|0\rangle\,.
\ee
The above expression conveys the idea that the ``theta'' vacuum is actually a condensate of $a$- and $\tilde{a}$-particles, which somehow resembles the situation we have already encountered with mixing. However, in order to construct an effective comparison with the latter framework, we need to reformulate Eq.~(\ref{thetavac2}) as
\be\label{thetavac3}
|0(\vartheta)\rangle=e^{-\frac{\mathcal{K}}{2}}e^{\sum_ka_k^\dagger\tilde{a}^\dagger_k}|0\rangle=e^{-\frac{\tilde{\mathcal{K}}}{2}}e^{\sum_ka_k^\dagger\tilde{a}^\dagger_k}|0\rangle\,,
\ee
where
\be\label{ent}
\mathcal{K}=-\sum_k\Bigl(a^\dagger_ka_k\ln\sinh^2\vartheta_k-a_ka^\dagger_k\ln\cosh^2\vartheta_k\Bigr)\,,
\ee
is called the entropy operator~\cite{tfd}, because its vacuum expectation value multiplied by $k_B$ yields precisely the entropy of the physical system. As a matter of fact, in order to make contact with a feasible physical picture and describe QFT for free fields at finite temperature, it is possible to prove that the arbitrary factors $\vartheta_k$ should satisfy the relation
\be\label{physics}
\beta\,\omega_k=-\ln\tanh^2\vartheta_k\,,
\ee
with $\beta=(k_BT)^{-1}$ being the
inverse of the emerging temperature $T$ of the
thermal vacuum $|0(\vartheta)\rangle$  and $\omega_k=\epsilon_k-\mu$, where $\mu$ is the chemical potential. 
Remarkably, in the limit $\theta_k\rightarrow0$ 
(i.e. for $T$ small enough), Eq.~\eqref{thetavac2} reads
\be
\label{neweq}
|0(\vartheta)\rangle\simeq\textbf{1}+\sum_ka^\dagger_k\tilde a^\dagger_k\vartheta_k|0\rangle\,.
\ee
As a final remark, we notice that the thermal 
vacuum~\eqref{thetavac2}, obtained by
augmenting the physical Fock space by a fictitious ``tilde''
space, has the same condensate structure of the 
vacuum perceived by stationary (uniformly accelerating) 
observers outside black holes  (in Minkowski spacetime).
Of course, in that case the 
dual space can be interpreted
in terms of the particle states
on the hidden side of the horizon~\cite{Israel, unruh}
and the entanglement arises among
modes across the horizon.

\end{document}